State-Level Representation of Chinese Scholars Mobility to and within the United States, 2009-2018

Caroline S. Wagner[1,2], Xiaojing Cai[3], and Jeroen Baas[4]

Abstract

A review of researcher mobility between China and the United States shows overall growth in mobility between the two nations. A review based upon a scholar's change of address from China to the United States or vice versa reveals mobility patterns. Between 2009 and 2018, an overall upward trend is noted for incoming scholars from China, as well as upward trends in those returning to China. Within the United States, researchers flow across states, but they are less likely to leave the larger, more R&D-rich states like California, New York, or Massachusetts. Chinese researchers are more likely to locate based upon reputation of research institutions than overall research spending or defense research locations. Mobility has a positive impact on China-U.S. collaboration numbers and citations.

---

[1] John Glenn College of Public Affairs, Ohio State University, Columbus, OH 43210 USA ; ORCiD: 0000-0002-1724-8489
[2] Corresponding author Email: wagner.911@osu.edu
[3] Department of Public Administration, Yangzhou University, Yangzhou 225127, China; and ORCiD: 0000-0001-7346-6029
[4] Elsevier B.V. Registered Office, Radarweg 29 1043 NX Amsterdam, The Netherlands and International Center for the Study of Research, Elsevier, Radarweg 29, Amsterdam, The Netherlands, ORCiD:

Introduction

The mobility of scholars has greatly increased over the past decades, but particularly so for Chinese nationals moving to the United States. Even in the face of visa challenges and political tensions, the number of international Chinese students and scholars moving to the U.S. for higher education and to participate in and conduct research and development (R&D) continues to grow. An estimated 4,400 Chinese scholars (which can include students) came to the United States in 2017, joining more than 9,000 who preceded them and who remain in the U.S. (Cao et al., 2019). Chinese students studying in the United States in 2017 numbered about 141,000 undergraduates, and 125,000 graduate students (NSB, 2019). Chinese doctorate earners graduating from U.S. universities in 2018 with plans to remain in the U.S., numbered more than 4,000—similar to the preceding five years (NSB, 2019).

An aggregate active number of Chinese scholars (not students) who joined U.S. laboratories in the United States is difficult to estimate but may be an inflow of about 5000. We are not interested in details of specific individual decisions to move abroad, since we expect these reasons to differ for each person. However, we are quite interested in group-level movements and what they can reveal about the geographic spread of foreign visitors in the U.S. These data can give us insight into the strengths and attractiveness of the U.S. system. This article seeks to fill in some of the gaps in understanding of where Chinese researchers move to and from the United States as well as within the United States. We expect to show that Chinese researchers seek to cooperate in U.S. regions with elite institutions and plentiful federal funding.

Benefits of Mobility

The global mobility of people is considered a positive trend. In science and technology (S&T) related fields, the mobility of skilled workers is recognized as an important aspect of innovation. The OECD has studied the movement of an educated class (2008, 2016), noting the contribution of mobility to the creation and diffusion of knowledge through direct interactions. We assume scholars move to consume some kind of resource that is available at the destination location. This 'resource' can be (and surely is) the intangible feature of 'prestige'; elite institutions have accumulated physical and intellectual infrastructure as well as well-known faculty. Most mobile scholars desire to attach to a prestigious person or institution; this results in some places being over-solicited. Those in control can choose to admit or welcome people who meet certain criteria, such as those coming from a similarly prestigious institution. From there, we can imagine a cascading effect of matching people with places based on some mix of prestige, reputation, resource availability, capabilities, and concentrations of knowledge.

The movement of talented researchers has been termed 'brain circulation', a phenomenon that is seen as accelerating development in both the visited and home countries (Saxenian, 2005). The benefits of circulation can only be gained if the home country have the absorptive capacity to use the knowledge gained by mobile scholars (Song, 1997). Without requisite absorptive capacity, the emigration of talented workers is a 'brain drain' for the sending country. As home countries increase their domestic R&D spending and S&T education, absorptive capacity grows and émigrés return home in larger numbers. This can be seen in the example of Korean and Taiwanese U.S.-based doctoral candidates who graduated in 1996: only 21% and 40%, respectively, were still working in the US in 2001, whereas 86% and 96% of the 1996 cohort of doctoral graduates from India and China were still working in the US in 2001 (Jonkers & Tijssen, 2008). Cao et al. (2019) showed that an increasing number of Chinese researchers are returning home in the late 2010s (Cao et al., 2019), although the number leaving is still balanced with inflows.

Gu et al. (2016) suggested that China is unable to fully absorb tertiary-level workers and this fact may dissuade some émigrés from returning.

Without requisite absorptive capacity, the emigration of talented workers is a 'brain drain' for the sending country (Gaillard & Gaillard, 1997). Scientific émigrés often collaborate with researchers in their native country, which helps to build capacity. When talented foreign visitors return home, it has been termed a 'reverse brain drain' (Jonkers & Tijssen, 2008). After returning home, however, many retain close ties to the nation they visited and continue to publish with foreign colleagues (Jonkers & Cruz-Castro, 2013). China has made the R&D investments to support endogenous innovation (Gu et al, 2016). Cao et al. (2019) showed that China's international collaborative work is much more highly cited than national work, suggesting that the focus of much of China's scientific research community is actively engaged at the international level, and a majority of these authors had been abroad.

Literature addresses the fact that a nation's knowledge is not contained within its physical borders, although the mode of transfer and 'spillover' is a source of debate. A long-standing discussion continues on what kinds of knowledge can be transferred as 'disembodied' from a person[5], versus how much knowledge is tacitly known (and carried) by individuals. This question remains unresolved. It is clear that many nations contribute knowledge to the work of other nations (Lee, 2006), although a measure is elusive. Economists have studied this question by defining and attempting to measure knowledge 'spillovers' (see Griliches, 1979) by assessing three contributing factors: import flows, cross-border investments, and disembodied direct channel. Guellec & van Pottelsberghe (2004) found that these disembodied cross-border spillovers make a significant contribution to knowledge transfer. Very little has been done to measure the contribution of the movement of people to national capacity, although Wagner & Jonkers (2017) showed that open countries have strong science.

Cowan and Jonard (2005) modeled mobility within cross-border knowledge transfer. They suggest that researcher mobility is embedded in complex dynamics:

> First, it permits a more global "standing on the shoulders of giants"—when an agent innovates, he is not re-inventing the wheel, and is, in fact starting from a higher level if he has had access to innovations from other parts of the world. Second, most innovation is the recombination of existing ideas. Thus increasing the general level of access to ideas, particularly if it increases as well the range of types of ideas that are easily assimilated should facilitate knowledge creation…. The parts of knowledge that are hard to transmit as codified knowledge are typically transmitted in close, (often face-to-face), interactions. (p. 187-188)

Using a Herfindahl index in a model of the knowledge system, Cowan & Jonard (2005) found that mobility contributes to knowledge growth. Using network tools, the model increased the links among agents; knowledge levels were measured under different network topologies. As permanent links are added to the model, the measurable knowledge levels increase for groups up to a certain size. (It is possible to have too much networking, they found.) Moreover, knowledge levels increased as the distribution of networking is more uniform across agents. They found: "bringing more people into the set of those who have long distance, permanent connections is valuable…" to a nation's knowledge base (p. 205). This suggests that welcoming foreign visitors, and retaining ties after visitors return to their country-of-origin, continues to strengthen both parties following a face-to-face visit.

---

[5] Guellec and van Pottelsberghe de la Potterie (2004) list different ways that 'disembodied knowledge' can be transferred between countries: The purchasing of patents, licenses or know-how from foreign firms, observe and copy competitive products (e.g. reverse engineering), hire foreign scientists and engineers, interact with foreign competitors who invested in their country (foreign direct investment), read the scientific and technological literature, and have direct contacts with foreign engineers in conferences or fairs (p. 357).

This article presents additional data to extends an analysis conducted of the movement of Chinese scholars (Cao et al., 2019) that showed Chinese scientists were more likely to come to the United States than the European Union, and also more likely to remain in the United States. This led to additional questions about where in the United States these Chinese scholars visit, and the further question of what they may be contributing to the U.S. knowledge base.

Data and Methods

To answer the question of which States receive the most Chinese scholarly visitors, a novel bibliometric approach is used here. The methodology was developed for and presented in Cao et al. (2019); it offers an approach to estimate the number of Chinese scholars in the United States. (These works also build on Wagner et al, 2018; Sugimoto et al, 2017; OECD, 2017; Moed and Plume, 2012). An analysis of the micro-data on publications contained in Elsevier's Scopus database allowed us to track researchers from the moment of first publication, whether in the United States or in China, and identify the fact that the author migrated within the USA or to another nation. It is thus possible to trace Chinese researchers who have first published in China and subsequently published in the United States (by state). The number of researchers who started their publishing career in China, followed by publications in the USA, and who then continued to publish at the time of the analysis are used as a proxy for mobility.

This measure has limitations. The method counts people who published in both China and the United States. Not counted are students of Chinese origin who earn a doctoral degree in the US and remain in the US after finishing their studies, numbering more than 4000, as noted above.

(In 2018, the National Science Board reported that of the 6,182 Chinese students earning doctorates in U.S. universities, 79.4% planned to stay, so our survey does not count about 4,900 Chinese citizens, for 2018, who will be working in the United States (NSB, 2019)). Many of these researchers will not have published in journals indexed in Scopus before moving to the US and therefore, they are not captured in our analysis. However the measure does offer an important improvement/alternative over current surveys. The method focuses on active scientists, excluding those overseas Chinese PhD holders who pursue a different career that does not involve publishing.

In order to develop the data on Chinese nationals within the U.S. states, mobility indicators were created at the author level and aggregated at US States level by year. Counts were derived from scholarly, refereed publications indexed in Scopus between 2009-2018. Active researchers are authors who publish articles in a given year. Inflow to a State is identified as a researcher who published in China and then newly published in a U.S. State. Conversely, outflow from a State is identified by a researcher who published in that State, but, the next year, published in China. (This can also include people who are neither American nor Chinese, but we have no way to identify these people. We expect this number to be small.) Based on the destinations and origins, the inflow (or outflow) is split into domestic and international inflow (or outflow), and the latter further split into inflow from (or to) China and inflow from (or to) other nations excluding China. (Here we'd like to note that inflow from China, or outflow to China are related to but not equal to Chinese students and scholars because they are not limited to Chinese nationals.) Outflow to China includes US researchers starting to work full-time or part-time in China, and inflow from China can include a third-country scholar who used to work in China and then takes on a new position in the US.

To assess the relative inflow and outflow, we examined two sets of relative indicators. First is the percentage share of Chinese inflow (or outflow) as a share of total inflow and outflow; this reveals the relative share of Chinese researchers compared to all the mobility activity of US States. The calculation of percentage share of Chinese inflow in year t is:

$$\%(\text{China inflow in Total}) = \frac{\text{\# inflow from China in year t}}{\text{\# total inflow in year t}} * 100\% \quad (1)$$

The other set of indicators are the (total, domestic, international or China) rate of inflow/outflow in active researchers in the previous year for a single State:

$$\text{Rate of total inflow} = \frac{\text{\# Total inflow in year t}}{\text{\# active researchers in (t}-1)} \tag{2}$$

$$\text{Rate of domestic inflow} = \frac{\text{\# Inflow from other States in year t}}{\text{\# active researchers in (t}-1)} \tag{3}$$

$$\text{Rate of international inflow} = \frac{\text{\# Inflow from other nations in year t}}{\text{\# active researchers in (t}-1)} \tag{4}$$

$$\text{Rate of China inflow} = \frac{\text{\# Inflow from China in year t}}{\text{\# active researchers in (t}-1)} \tag{5}$$

The above set of indicators was calculated for 20 states with the largest number of active researchers and for two years (2010 and 2018).

To assess the reasons for the attractiveness of different States, we gathered the amount of federal R&D funding from the National Science Board (NSB)[6] and the number of research practitioners by State and by year. Two indicators were chosen to examine the relationship with inflow or outflow. First, we calculated the percentage of R&D allocated to each State as a percentage of US total R&D. Second, we calculated the federal funding-per-researcher by dividing the funding by the number of active researchers. We analyzed the Chinese presence based upon the characteristics of U.S. States.

We added citation data to assess the relationship between a State's scholarly impact and its role in attracting Chinese visitors. To compare Chinese mobility to citation impact at the State level, we use a measure that provides a fractional allocation of credit based upon the State appearing in the address of authors. The measure is called the fractional Field-Weighted Citation Index (fractional FWCI) defined by Scopus, which was previously used in Wagner et al. (2018). FWCI is a normalized citation impact measure, defined as ratios of observed citations to the expected citations for articles in same document type, same subject category and same publishing year. Here, fractional FWCI is calculated as

$$\text{Fractional FWCI} = \sum(f_i \text{FWCI}_i)/\sum f_i \tag{6}$$

Where $f_i$ is the proportion of authors of article i belonging to a State.

---

Results

As shown in Figure 1, between 2009 and 2018, researchers moving from China to the United States, and the number moving from the USA to China increased. The numbers of both types of flow has more than doubled in past 10 years, from about 2,500 (total) in 2009 to about 5,000 (total) in 2018. The number of the outflow and inflow are almost equal, indicating a balanced flow from China to US and those returning.

Figure 1. Intensity of inflow from China to the US and outflow from US states to China, 2009-2018.

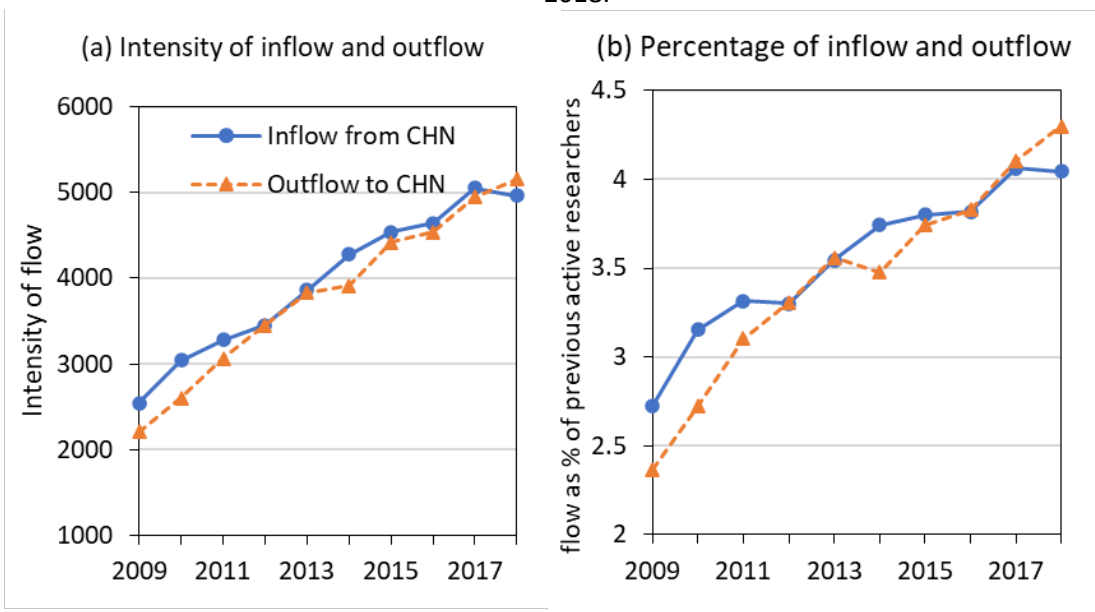

(a) Intensity of inflow and outflow

(b) Percentage of inflow and outflow

For the 50 states, when accounting for years 2010 and 2018, the rate of total inflow (including inflows of US and foreign researchers) per year from China ranged from 10-14% of all inflow. No significant increase is seen as a percent of inflows from 2010 to 2018. Compared to 2010, states like Maryland, Tennessee, Washington, Colorado and Massachusetts showed a lower rate of total inflow in 2018 than in 2010; only Florida showed a slight increase.

A breakdown of flows by State is shown in Figure 2a (inflows) and 2b (outflows) comparing 2010 with 2018. The rate of total outflow of previously active researchers ranged from 11-14.4% in 2010 and 10-13.8% in 2018, indicating a drop of relative outflow for 20 states from 2010 to 2018. The domestic inflow or outflow, which was the main component of the total inflow or outflow, accounted for about 8.5% of total inflow or outflow on average. Massachusetts showed highest rate of international flows for both inflow and outflow. From 2010 to 2018, only the states of New Jersey, North Carolina, and Florida increased in rate of international inflow. In Figure 2b, almost all states show lower rate of international outflow in 2018, especially Maryland, California, Florida, Indiana, Colorado and Ohio. China inflow and outflow were the only exception with higher rate in 2018.

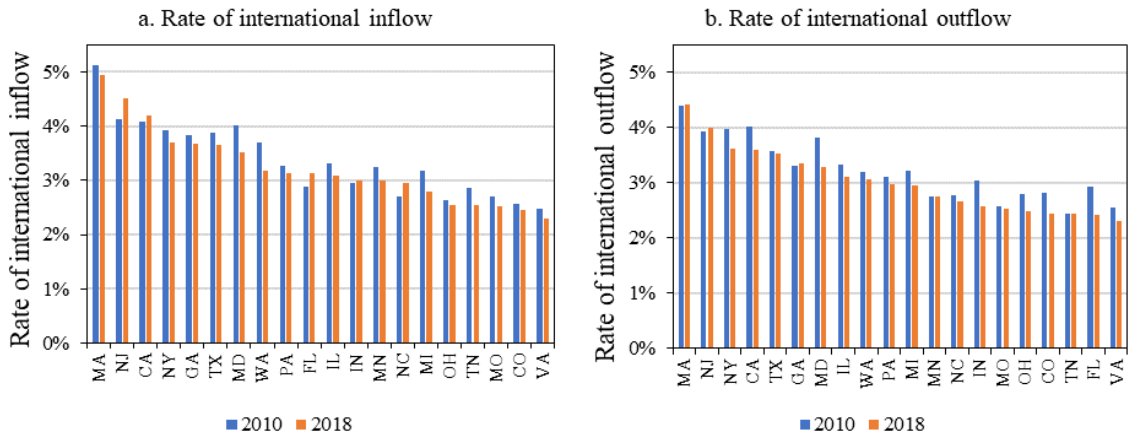

Figure 2. Rate of international inflow (a) and outflow (b) of previous active researchers for 20 states, 2010 and 2018.

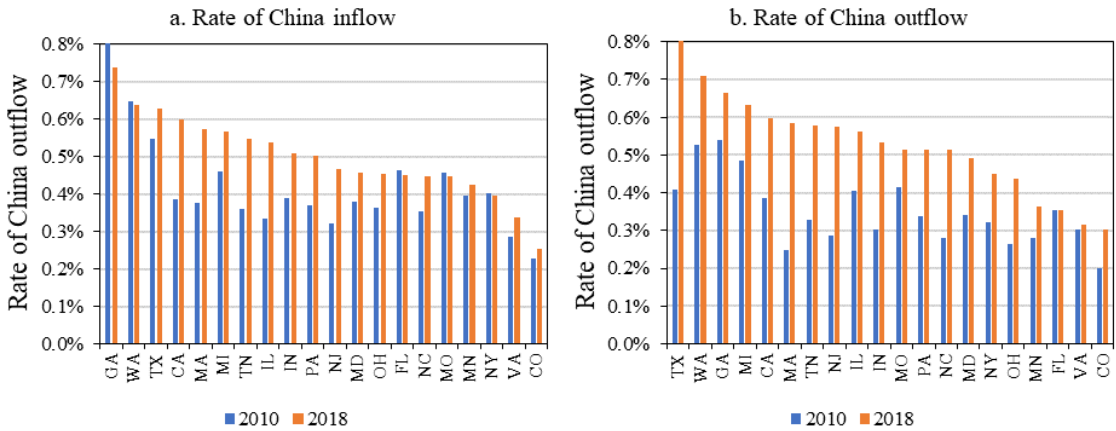

Figure 3 shows the rates of China inflow and outflow in 2010 and 2018 for 20 states. As shown in

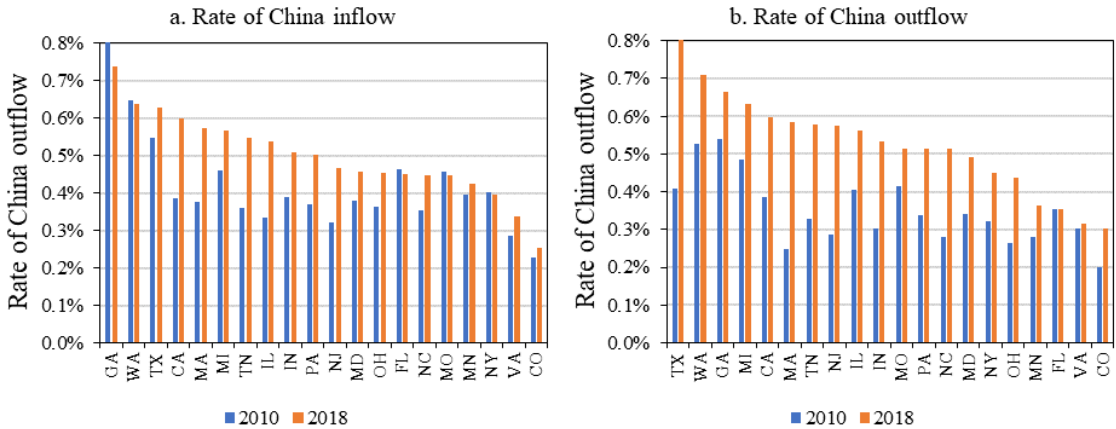

Figure 3a, the rate of Chinese scholars flowing into a State ranged from 2.5% of inflows in Virginia to 5% for Massachusetts. Most states, like California, Texas, Connecticut, Massachusetts, Michigan, Tennessee, Illinois, Indiana, and New Jersey, showed clearly higher rates of China inflow in 2018. However, Georgia and Washington, with the highest rate of China inflow (0.75% and 0.65%), did not show higher rate in 2018. In

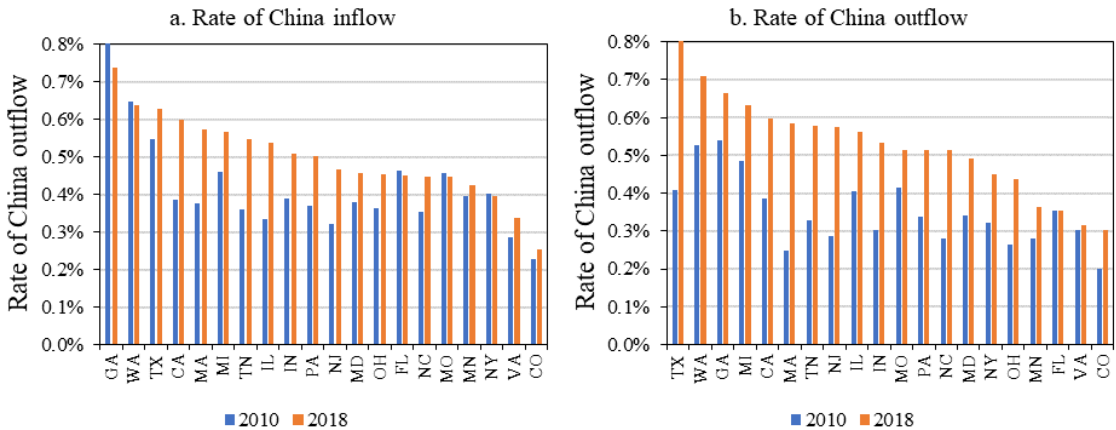

Figure 3b, Texas showed the highest rate of China outflow in 2018--the most dramatic increase during the period. Other states also increased their rate of China outflow in 2018, except for Florida and Virginia. The results suggest an increasing role of China in the global network, with more intensive migration between China and most US States.

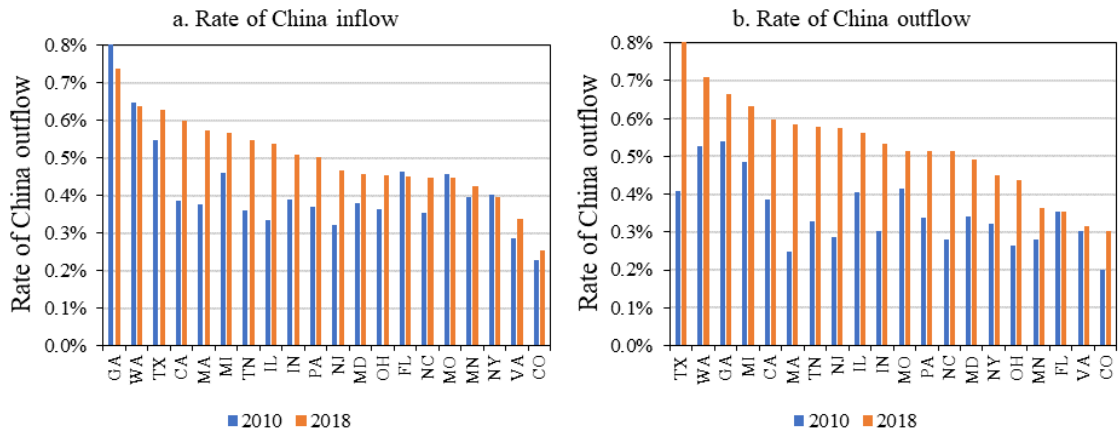

Figure 3. Rate of China inflow (a) and outflow (b) of previous active researchers for 20 states, 2010 and 2018.

Which states do Mobile Scientists related to China Move to and From?

Table 1 presents the number of actively mobile researchers in the United States by top-20 State. According to the number of active researchers in each state, California is the largest state with most researchers and largest flows (in and out) of researchers. New York is the second largest state, followed by Massachusetts, Texas, Pennsylvania, and Maryland.

Table 1. Basic Profiles of Mobility by State, 2018

| State | # Active Researchers | Inflows of Scholars | | | Outflows of Scholars | | |
|---|---|---|---|---|---|---|---|
| | | Total # of all inflows | Inflows from China | % of all inflows originating in China | Total # of all outflows | Outflows to China | % of all outflows leaving for China |
| CA | 129,361 | 14,787 | 739 | 5.00% | 12,372 | 740 | 5.98% |
| NY | 81,112 | 8,806 | 310 | 3.52% | 9,044 | 352 | 3.90% |
| MA | 66,624 | 8,251 | 372 | 4.51% | 7,855 | 378 | 4.81% |
| TX | 62,108 | 6,665 | 377 | 5.65% | 6,553 | 482 | 7.35% |
| PA | 49,390 | 5,409 | 243 | 4.50% | 5,828 | 250 | 4.28% |
| MD | 47,202 | 6,000 | 215 | 3.58% | 6,048 | 230 | 3.81% |
| IL | 38,072 | 4,102 | 201 | 4.90% | 4,686 | 210 | 4.49% |
| OH | 34,515 | 3,357 | 153 | 4.54% | 3,642 | 147 | 4.04% |
| FL | 32,645 | 3,725 | 141 | 3.77% | 3,467 | 110 | 3.19% |
| NC | 32,056 | 3,510 | 138 | 3.93% | 3,472 | 158 | 4.55% |
| MI | 31,450 | 3,082 | 172 | 5.58% | 3,362 | 192 | 5.70% |
| GA | 25,908 | 2,863 | 185 | 6.45% | 2,855 | 167 | 5.83% |
| VA | 24,216 | 3,191 | 81 | 2.55% | 3,141 | 76 | 2.42% |
| WA | 22,649 | 2,809 | 143 | 5.10% | 2,601 | 159 | 6.12% |
| NJ | 21,016 | 2,923 | 99 | 3.40% | 2,725 | 123 | 4.51% |
| IN | 21,005 | 2,127 | 104 | 4.90% | 2,224 | 109 | 4.88% |
| MN | 19,194 | 2,099 | 77 | 3.68% | 2,032 | 66 | 3.24% |
| CO | 18,980 | 2,268 | 45 | 1.99% | 2,081 | 54 | 2.59% |
| TN | 17,166 | 1,786 | 90 | 5.05% | 1,879 | 95 | 5.07% |
| MO | 15,414 | 1,748 | 69 | 3.93% | 1,971 | 79 | 4.03% |

Note: Outflows that are defined as 'retired' are excluded when calculating the total outflow.

Figure 4 and Figure 5 provide a clearer view of which states have the most flowing scientists from China during 2009-2018, as a complementary to data shown in Table 1. The maps for outflow to China from US states are similar as the maps for inflow and are not shown here.

As shown in Figure 4, the States of California, Texas, New York, Massachusetts, and Pennsylvania are the most attractive states in terms of flowing scientists to and from China. A different group of states, i.e., Georgia, Nebraska, Texas, Oklahoma, Louisiana, Michigan, etc., show high percentages of flow related to China. States like California, in Figure 5, States of New York, and Massachusetts show relatively lower mobility flow as a percentage compared to the ranking in absolute terms. Texas is the only state showing a close relationship with China in both absolute and relative terms; California presented a higher percentage of inflow or outflow represented by China in recent years (Table 1). Georgia and Michigan are also high in terms of the absolute number of Chinese inflow and pretty high in terms of the relative share of Chinese inflow.

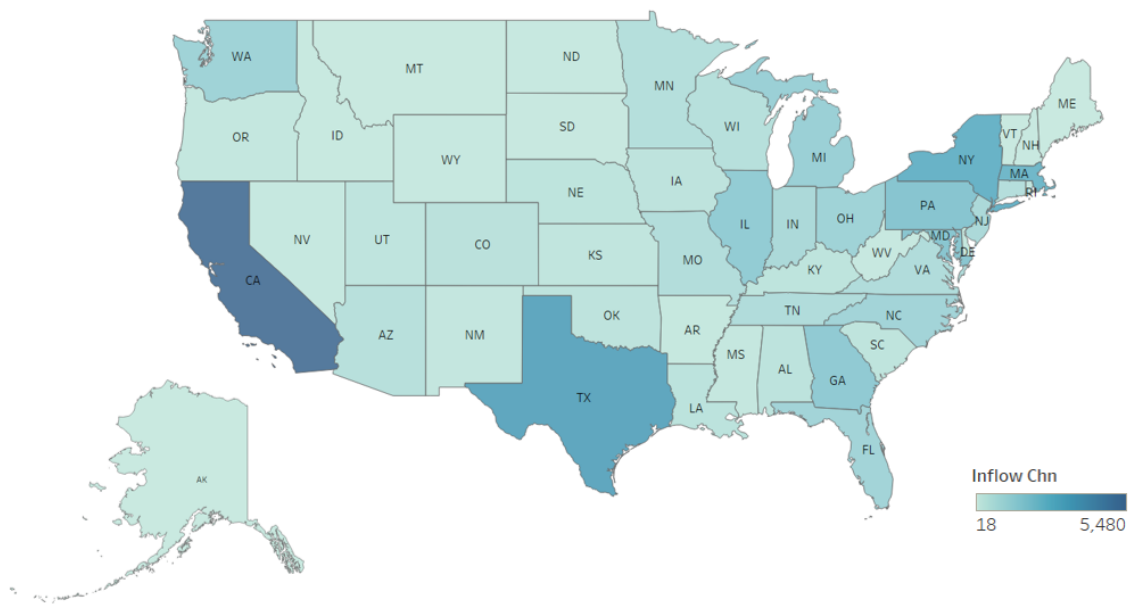

Figure 4. Intensity of inflow from China to the US by state, 2009-2018.

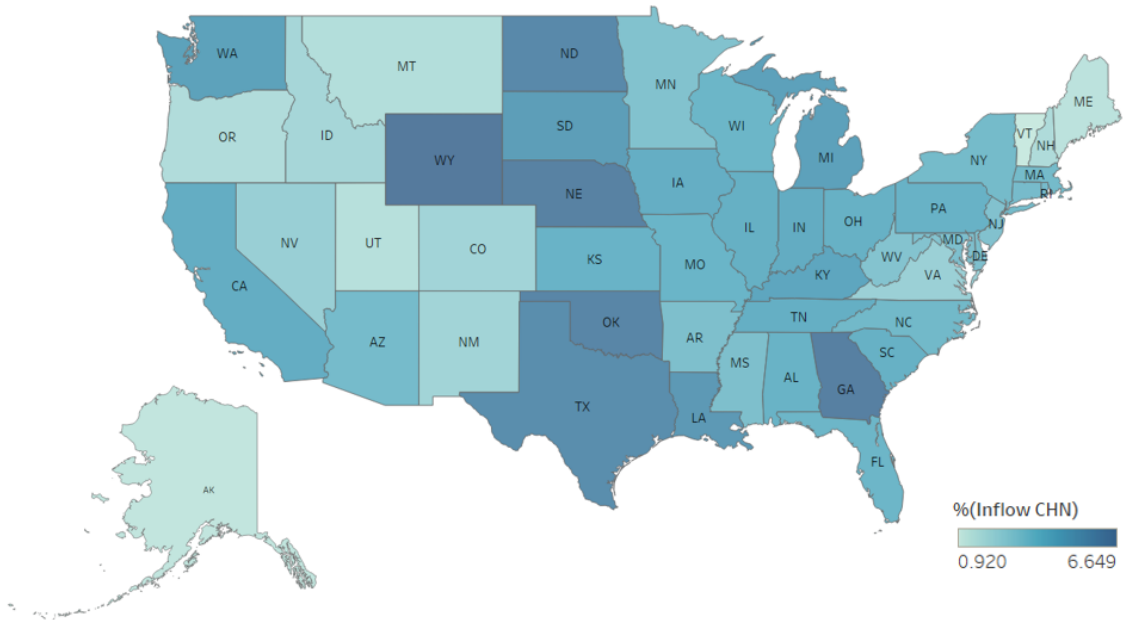

Figure 5. Percentage share of inflow represented by China, by state, 2009-2018.

R&D Expenditure and Mobility

Figure 6a shows that, in 2018, California—the largest State in R&D spending—attracted the largest number of Chinese researchers. In contrast, Maryland has high R&D spending but it did not attract the same number of Chinese researchers proportional to California, likely because it has a higher percentage of defense research activity which limits the ability of Chinese researchers to participate. After California, the States of Texas, Massachusetts, New York, Pennsylvania and Illinois attract larger numbers of Chinese researchers in descending order. The States with less attraction to Chinese visitors than expected are Virginia, Alabama, Colorado and New Mexico—we note that these States tend to have higher defense R&D spending—and thus have fewer Chinese researchers flowing through their States. Figure 6b shows the outflow of Chinese researchers leaving the United States, and the list is closely correlated to the inflow. The inflow and outflow are balanced at the State level.

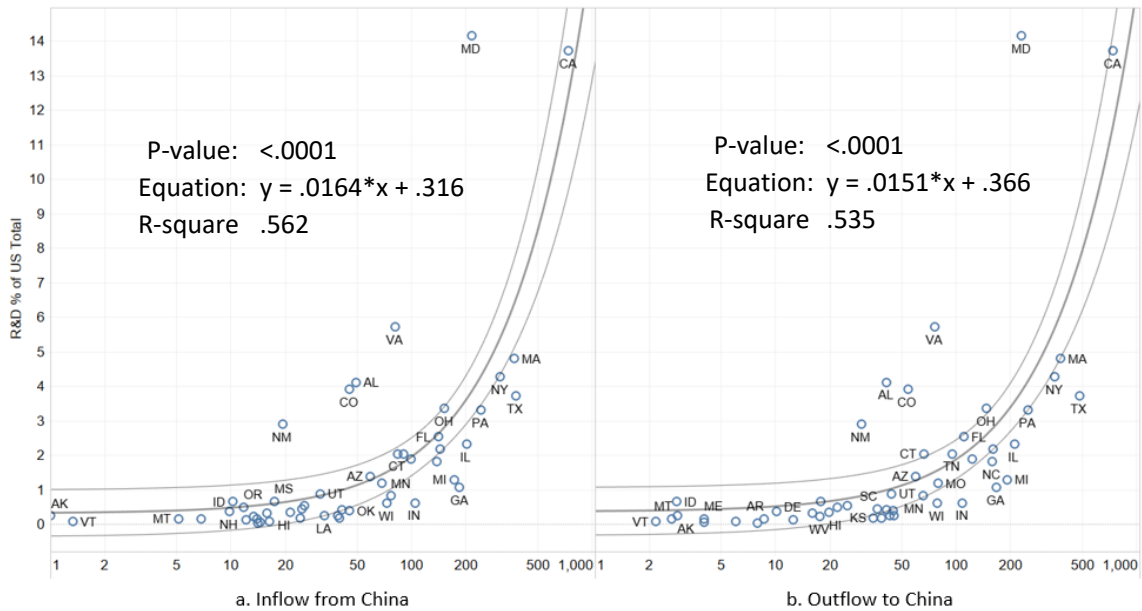

Figure 6. Intensity of inflow from China (a) and outflow to China (b) at State level in relationship to R&D spending, as % in US total, for 50 States, 2018.

When we examined federal R&D spending by number of active researchers (funding-per-researcher), we find no correlation between the attractiveness of a State and their funding-per-researcher. The calculation of funding-per-researcher reveals a different ranking of States from total spending. When calculating by funding-per-researcher, the top-ranked States are Alabama, New Mexico, Maryland, Virginia, and Idaho. This feature does not make a State more attractive to Chinese researchers. In fact, as shown in Figure 7, States that have relatively smaller amounts of funding-per-researcher remain highly attractive to Chinese researchers. California falls in the middle position in terms of funding-per-researcher, but it attracts the largest number of Chinese researchers. Similarly, most states that fall in middle bottom of Figure 7a, like Texas, Indiana, Washington, Michigan, and Massachusetts, have lower R&D funding-per-researcher, but show higher level of attractiveness to Chinese researchers than those with higher funding-per-researcher.

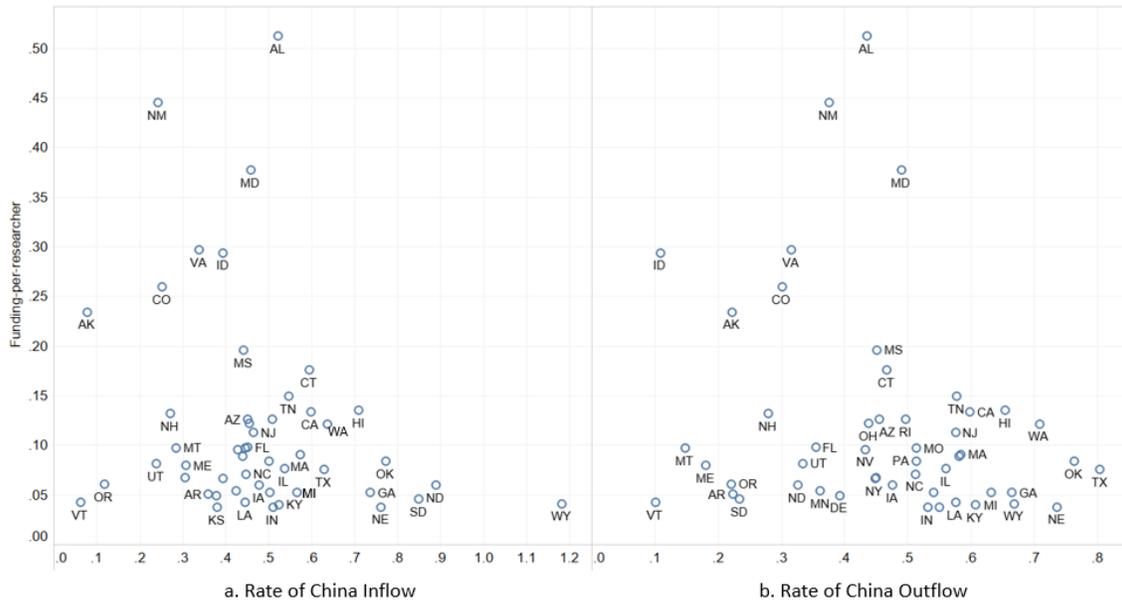

Figure 7. Rate of China inflow (a) and outflow (b) at state level in relationship to funding-per-researcher, for 50 states, 2018.

Mobility Flow, Collaboration and Citation Impact

The negative finding of attractiveness for States with higher funding-per-researcher suggests that Chinese researchers are seeking an intangible value when choosing a location in the United States; this value is likely the prestige of the institution they are visiting. A measure that reflects prestige is the number of citations garnered by institutions in a State. We aggregated citations at the State level to test whether Chinese visitors are attracted to States where institutions garner higher citations. Figure 8 a and b show the inflow and outflow of Chinese scholars in 2018 correlated to the fractional FWCI gained by international collaborations. We know from earlier research that Chinese-U.S. collaborative publications are more highly cited than U.S. or Chinese publications alone (Wagner et al. 2015).

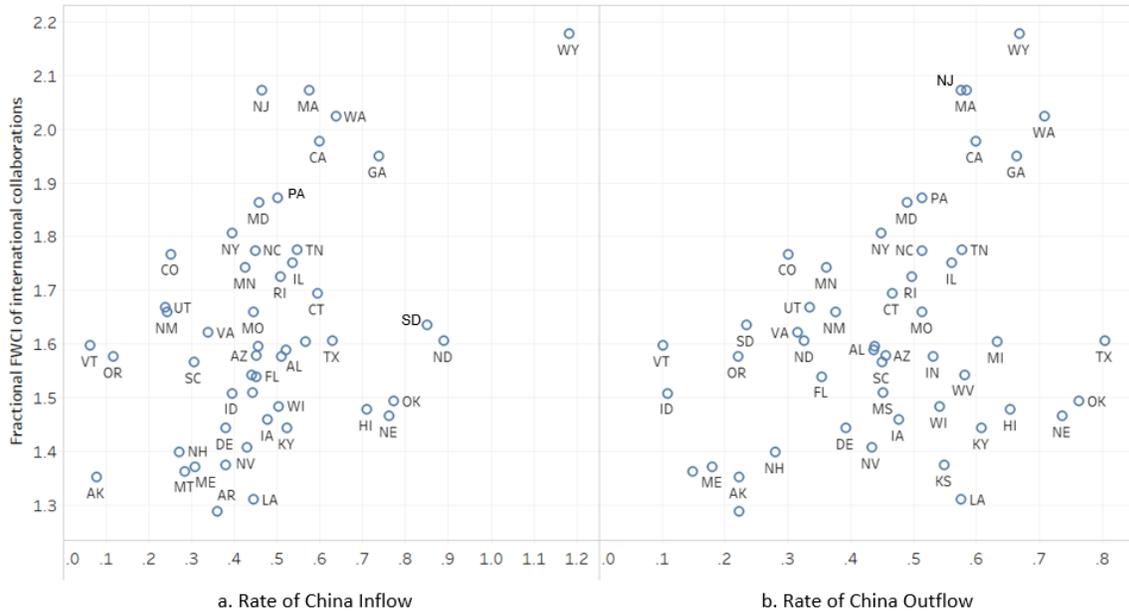

Figure 8. Rate of China inflow and outflow in relationship to fractional FWCI of international collaborations, for 50 states, 2018.

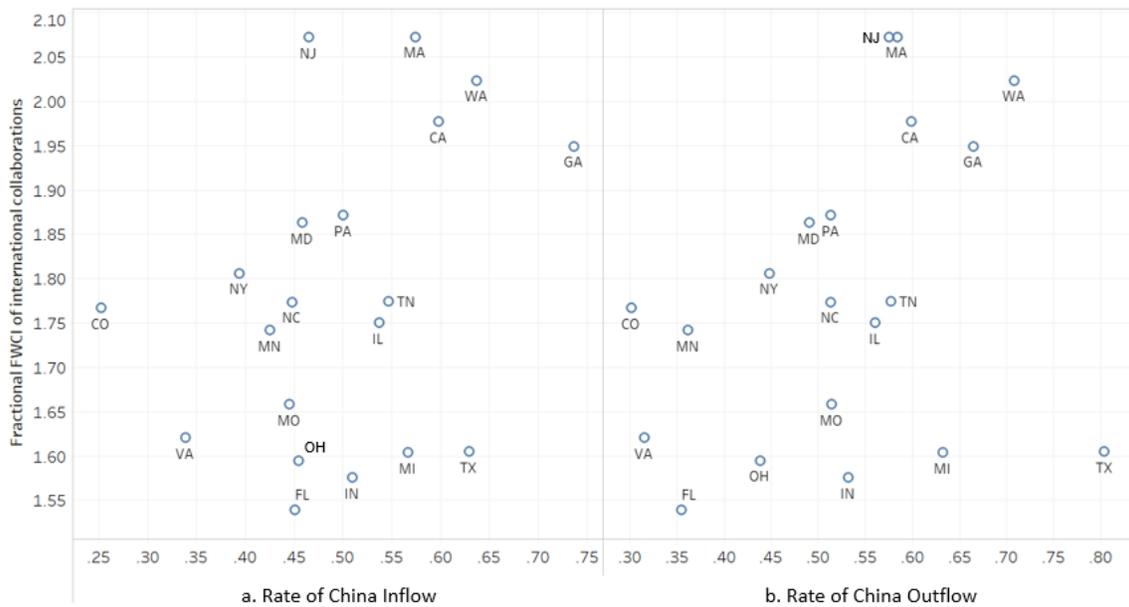

Alternative: Rate of China inflow and outflow in relationship to fractional FWCI of international collaborations, for 20 states, 2018.

Figure 8 shows the close correlation between citation strength and the flow of Chinese researchers. As expected, we see a very close relationship between the prestige factor of a State and its attractiveness to Chinese researchers in Figure 8a. In 2018, international collaborations of Massachusetts and New Jersey gain the highest citation impact, followed by Washington,

California, and Georgia. These high-impact states show consistent high rate of China inflow except for New Jersey that is not as attractive to researchers from China. States of Virginia and Colorado that show lowest ratio of China inflow gain lower citation impact than most states. Exceptions are Texas and Michigan, which are attractive to Chinese researchers but exhibit quite low citation impact. Figure 8b shows similar case in terms of ratio of China outflow and citation impact.

Table 2 shows a fixed effect test to seek further confirmation of the relationship between Chinese migration and citations. The regression shows that, with holding other variables constant, the rate of China inflow was positively associated with higher fractional field weighted citation index numbers (FWCI) of both total articles and articles involving international collaboration, indicating a positive link between Chinese inflow and state citation impact. Increasing ratio of China outflow do not impact citation rates significantly.

Table 2. Two-way fixed effects models of Chinese inflow/outflow on citation impact, for 50 states, 2009-2018.

|  | (1) Frac FWCI | (2) Frac FWCI Intl |
|---|---|---|
| Ratio of China inflow | $0.108^{*}$ | $0.362^{**}$ |
|  | (0.056) | (0.159) |
| Ratio of China outflow | 0.031 | -0.027 |
|  | (0.047) | (0.113) |
| Ratio of other inflow | 0.0004 | $-0.001^{***}$ |
|  | (0.0003) | (0.0003) |
| Ratio of other outflow | $0.005^{***}$ | $0.007^{***}$ |
|  | (0.001) | (0.002) |
| State fixed effects | Yes | Yes |
| Year fixed effects | Yes | Yes |
| Constant | $1.180^{***}$ | $1.597^{***}$ |
|  | (0.025) | (0.059) |
| Observations | 486 | 486 |
| r2 | 0.262 | 0.093 |
| F | 8446 | 63.02 |

Robust standard errors in parentheses

$^{*}$ $p < 0.1$, $^{**}$ $p < 0.05$, $^{***}$ $p < 0.01$

We further show that high ratios of outflow to China and inflow from China correspond to extensive China-US collaboration (Figure 9). Mobile scientists who flow across national borders, regardless of the direction of flow, tend to collaborate with scientists from both countries, playing a bridging role to connect them.

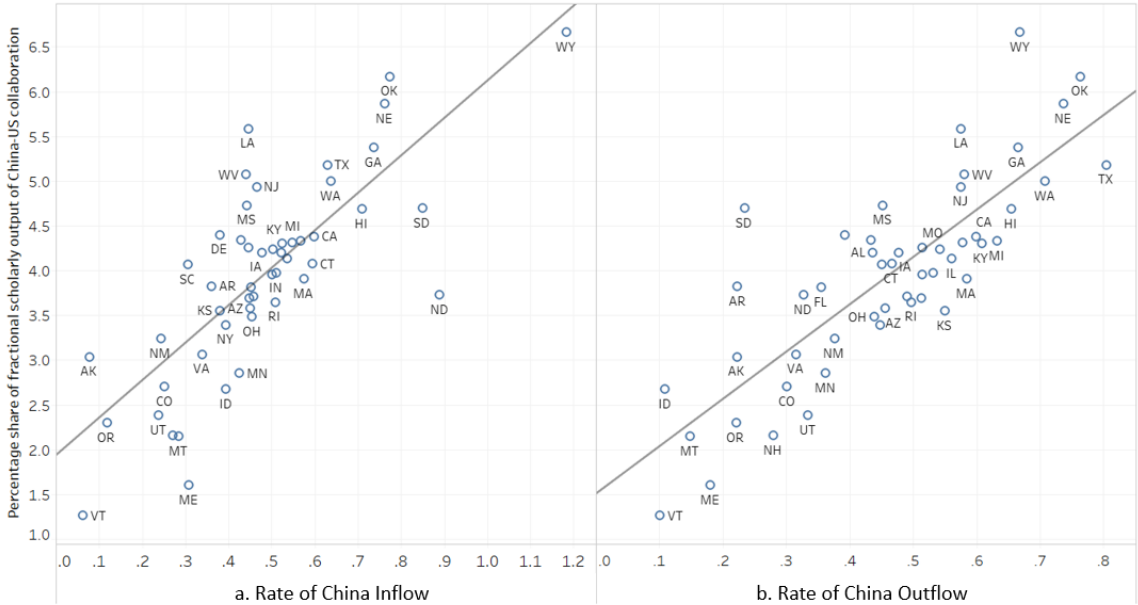

Figure 9. Ratios of China inflow (a) and outflow (b) in relationship to percentage share of China-US collaboration, for 50 states, 2018.

Table 3 confirms the above finding about Chinese mobility and international collaboration. Model 1 shows that higher ratio of outflow to China is associated with higher number and share of China-US collaboration. This suggests that relationships formed in person may be retained and continued at a distance once a Chinese scholar returns to China. When using the percentage share of China-US collaboration in total number of articles of each state as dependent variable, the result indicated that both ratio of inflow from China and ratio of China outflow are positively related to higher share of China-US collaboration.

Table 3. Two-way fixed effects models of Chinese inflow/outflow on China-US collaboration, for 50 states, 2009-2018.

|  | (1)<br>$\#$ of intl_coCHN | (2)<br>$\%$ of intl_coCHN |
| --- | --- | --- |
| Ratio of China inflow | 101.765<br>(75.625) | 0.005$^{*}$<br>(0.003) |
| Ratio of China outflow | 184.818$^{**}$ | 0.012$^{***}$ |

|  | (1) | (2) |
|---|---|---|
|  | # of intl_coCHN | % of intl_coCHN |
|  | (77.340) | (0.002) |
| State fixed effects | Yes | Yes |
| Year fixed effects | Yes | Yes |
| Constant | 53.775 | 0.012*** |
|  | (60.703) | (0.001) |
| Observations | 486 | 486 |
| r2 | 0.364 | 0.534 |
| F | 4.248 | 53.249 |

Robust standard errors in parentheses

* $p < 0.1$, ** $p < 0.05$, *** $p < 0.01$

The results of Table 2 and Table 3 show that more researchers coming from China to the US increase the absolute and relative frequency of bilateral cooperation (which may result from Chinese nationals in the United States working with colleagues in their home country), while it is the inflow of elite researchers from China rather than outflow of researchers to China that reflect the choice to locate in higher prestige states. Researchers flowing out to China contribute more to building linkages between the two nations, while researchers are more beneficial in increasing research capacity of states in the US.

Conclusion

Data was collected to reveal the inflow to and outflow of Chinese scholars in the United States for the years 2009-2018. The analysis shows an increasing number of researchers flowing between the United States and China in both absolute and relative terms in the 10 years studied. The number of researchers arriving in the United States from China has grown, shown in this paper as an increasing number of inflows to the States. Nevertheless, we see a growing number of outflows of Chinese researchers going back to China, as well. Other inflow or outflow data does not show a growing trend across states (total flow, domestic flow, and international flow) over the ten years studied.

Variations among states are shown in the number of Chinese scholars flowing in and out. States with the largest number of Chinese scholars (California, New York, and Massachusetts) show lower inflows and outflows, suggesting greater stability for scholars visiting there. States like

Georgia, Nebraska, Texas, Oklahoma, and Louisiana showed greater flows in and out. Texas showed the highest flow rate of Chinese scholars in and out of the state.

States claiming greater amounts of R&D funding show higher numbers of Chinese inflow and outflow, although we could not find a correlation between R&D spending per researcher and the intensity of Chinese inflow and outflow. We tested and did not find any correlation between the chosen location of Chinese scholars and defense-related research. Moreover, our tests showed that Chinese scholars are more likely to be located in places with elite research institutions than any other factor. This suggests that institutional reputation is the greatest attractor of Chinese visitors.

We find a positive relationship between Chinese mobility and the number of China-US collaborations. Several factors contribute to this phenomenon. One reflects that researchers visiting for an extended period may use a U.S. address on an article for work conducted collaboratively with a former colleague from China; this appears as a China-USA collaboration. Secondly, many scholars who visit the USA continue to cooperate with U.S. colleagues after returning home, contributing to greater collaborations. Thirdly, researchers visiting the U.S. may introduce a U.S. researcher to someone from China, initiating a collaboration that might not have occurred otherwise. Mobility plays a vital bridging role between the two nations, fostering collaboration and communication between Chinese and US scientific communities.